\documentclass[preprintnumbers,prd,]{revtex4}%
\usepackage[ansinew]{inputenc}
\usepackage{color}
\usepackage{hyperref}
\usepackage{verbatim}
\usepackage{subfigure}
\usepackage{amsmath}
\usepackage{float}
\usepackage{fancyhdr}
\usepackage{graphicx}
\usepackage{amsfonts}
\usepackage{amssymb}
\usepackage{float}

\providecommand{\U}[1]{\protect\rule{.1in}{.1in}}
\begin{document}
\title{Modeling an Aquifer: Numerical Solution to the Groundwater Flow Equation}
\author{V. V\'azquez-B\'aez}
\email{manuel.vazquez@correo.buap.mx}
\affiliation{Benem\'erita Universidad Aut\'onoma de Puebla, Facultad de Ingenier\'{i}a, 72570 Puebla, M\'exico.}

\author{A. Rubio-Arellano}
\email{ana.rubio@alumno.buap.mx}
\affiliation{Benem\'erita Universidad Aut\'onoma de Puebla, Facultad de Ingenier\'{i}a, 72570 Puebla, M\'exico.}

\author{D. Garc\'{i}a-Toral}
\email{dolores@ifuap.buap.mx}
\affiliation{Benem\'erita Universidad Aut\'onoma de Puebla, Facultad de Ingenier\'{i}a Qu\'{i}mica, 72570 Puebla, M\'exico.}

\author{I. Rodr\'{i}guez Mora}
\email{isrrael.rodriguez@correo.buap.mx}
\affiliation{Benem\'erita Universidad Aut\'onoma de Puebla, Facultad de Ingenier\'{i}a, 72570 Puebla, M\'exico.}

\begin{abstract}

We present a model of groundwater dynamics under stationary flow and governed by Darcy's Law of water motion through porous media, we apply it to study a 2D aquifer with water table of constant slope comprised of an homogeneous and isotropic media, the more realistic case of an homogeneous anisotropic soil is also considered. Taking into account some geophysical parameters we develop a computational routine, in the Finite Difference Method, that solves the resulting elliptic partial equation, both in a homogeneous isotropic and homogeneous anisotropic media. After calibration of the numerical model, this routine is used to begin a study of the Ayamonte-Huelva aquifer in Spain, a modest analysis of the system is given, we compute the average discharge vector as well as its root mean square as a first predictive approximation of the flux in this system, providing us a signal of the location of best exploitation; long term goal is to develop a complete computational tool for the analysis of groundwater dynamics.

\end{abstract}

\maketitle

\section{Introduction}

One of the most powerful tools to advance theoretical and practical knowledge in the characterization of water flow through ground porous media are computational numerical models, which must always be compared and calibrated with experimental measurements \cite{wang,fowler}. In the practice it is common to carry out these  kind of studies with a wide combination of geophysical methods \cite{lochbuhler}. Physical models are often given in terms of partial differential equations, which for the specific case of groundwater dynamics they turn out to be analytically unsolvable when one tries to find a solution over realistic domains and conditions. The numerical solution, in the bast majority of the occasions, is used to predict the adequate management of hydric resources as can be seen in \cite{comision}, where they present a numerical study of the Puebla Valley Aquifer in Puebla, Mexico.

Although actual aquifers are inhomogeneous and anisotropic, in order to construct a model, they are usually decompose in a great number of elements since it is feasible to solve certain problems assuming that each element behaves in its neighborhood as a small ideal aquifer \cite{bear}. In past years, there has been a lot of activity towards enhancing the way we model ground heterogeneity using geostatistical data and digital models or by the implementation of genetic algorithms \cite{marsily}. Additionaly, there has been intensive work in the line of finding analytical solutions via new definitions of the derivative operator with fractional order, like the Caputo-Fabrizio derivative, these kind of outlines take into account the aquifer heterogeneity characterizing it as a scale problem. See for example \cite{atangana:2016}, where they modify the groundwater flow equation replacing the conventional time derivative appearing in the transient term by the fractional derivative, thus allowing the description of material diffusion at different scales.

This paper is organized as follows, in section \ref{sec_iterativeformulation} we give a quick review of the formulation of the Finite Difference Method and its numerical implementation, we describe briefly the key ideas of three of the main iterative schemes of solution. In section \ref{sec_model} we present the groundwater flow model, based on Darcy's Law, which is a good approximation to the dynamics of fluids in porous media provided they fulfill certain conditions. In section \ref{sec_numerical} we present the results of implementing the iterative method to this problem and a brief discussion is given. Finally, in section \ref{conclusions} we give final remarks and conclusions along with future work perspectives.


\section{Finite Difference Method and the Iterative Formulation}
\label{sec_iterativeformulation}

\begin{figure}[h]
\centering
\includegraphics[height=7cm,width=11cm]{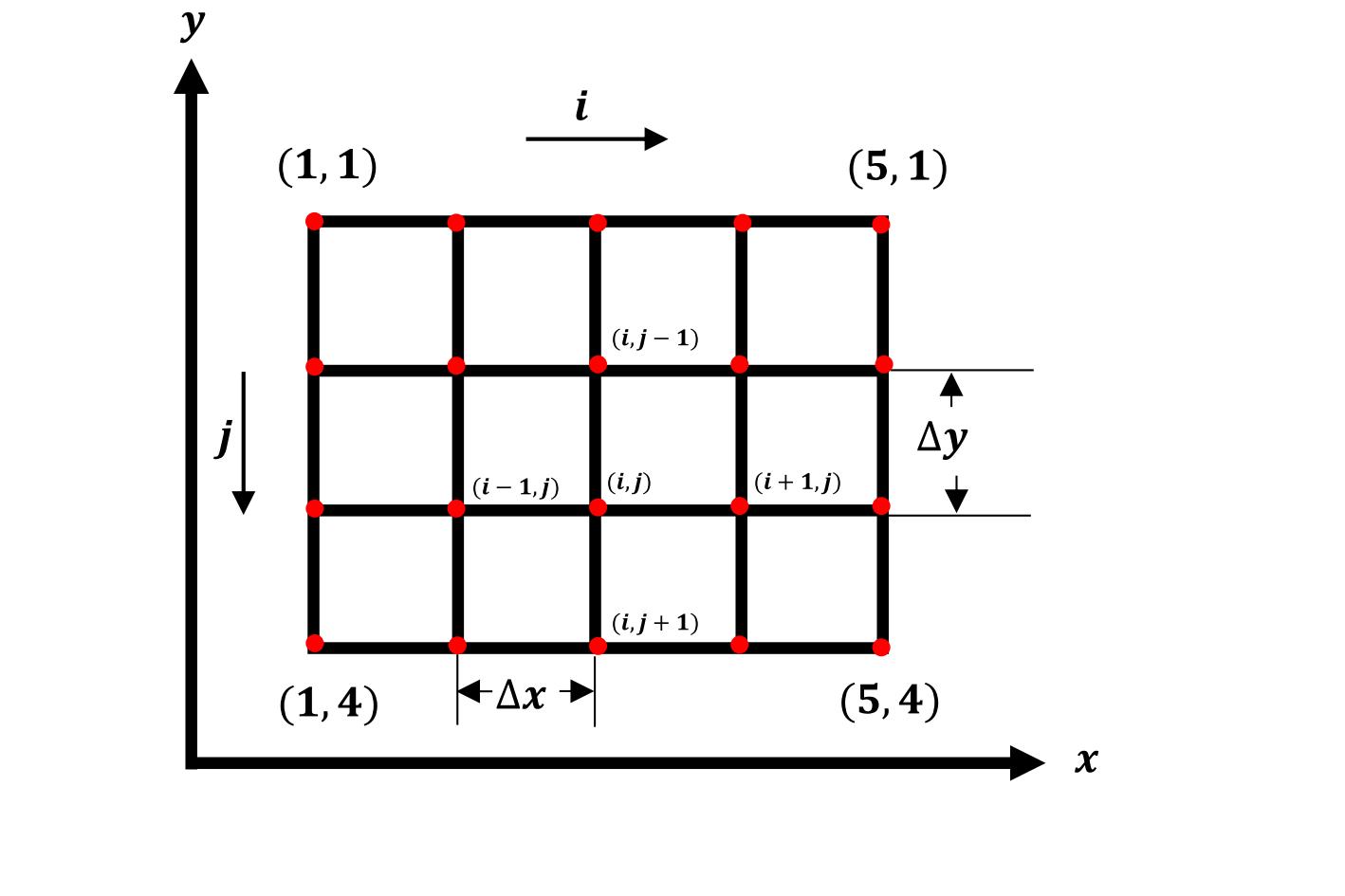}
\caption{{\protect\footnotesize {Index distribution of a point on the grid with step size of $\Delta x$ and $\Delta y$ over each coordinate direction.}}}%
\label{figgrid}
\end{figure}

The Finite Diference Method essentially transforms a differential equation into a system of algebraic equations by means of a spatial discretization of a physical problem's domain. This is performed considering a finite set of points on a rectangle, called grid, as shown in figure \ref{figgrid}.

Let $h(x,y)$ be a dependent variable for which we must solve the problem, we will use the central points scheme, so we fix our atention into the point labeled with indexes $(i,j)$ i.e., $h_{i,j}$. Each index labeled point is called a node, so that in the FDM the derivatives are approximated using the Taylor series and computed in terms of nodal points.

\begin{figure}[h]
\centering
\includegraphics[height=7cm,width=11cm]{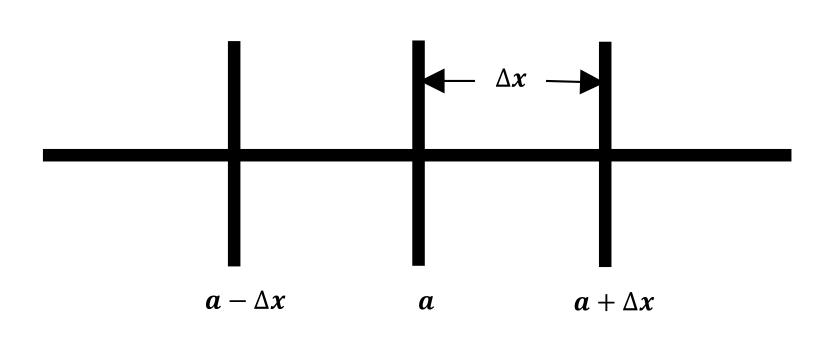}
\caption{{\protect\footnotesize {Interval centered at $x=a$ with a width equal to $2\Delta x$.}}}%
\label{figinterval}
\end{figure}

In order to show how this approximation is done let us focus on a single variable function $f$, so according to figure \ref{figinterval} if we Taylor expand this function to second order at points $x=a+\Delta x$ and $x=a-\Delta x$ we get
\begin{align*}
f(a+\Delta x)  & =f(a)+f^{\prime}(a)\Delta x+\frac{1}{2}f^{\prime\prime}(a)\left(  \Delta x\right)  ^{2}+\mathcal{O}(3),\\
f(a-\Delta x)  & =f(a)-f^{\prime}(a)\Delta x+\frac{1}{2}f^{\prime\prime}(a)\left(  \Delta x\right)  ^{2}+\mathcal{O}(3),
\end{align*}
substracting this two expressions and solving for $f^{\prime}$, ignoring higher order terms, we get
\begin{equation}
\label{firstderivative}
f^{\prime}(a)\approx \frac{f(a+\Delta x)-f(a-\Delta x)}{2\Delta x},
\end{equation}
where it is obvious the reason for the name central points scheme (or central differences scheme) for the procedure, the quantities for which one requires to solve are given in terms of the extreme points at each side. With this, equation (\ref{firstderivative}) allows us to approximate numerically the first partial derivatives as
\begin{equation}
\label{xpartial}
\frac{\partial h}{\partial x}=\frac{h_{i+1,j}-h_{i-1,j}}{2\Delta x},%
\end{equation}%
\begin{equation}
\label{ypartial}
\frac{\partial h}{\partial y}=\frac{h_{i,j+1}-h_{i,j-1}}{2\Delta y}.%
\end{equation}

Similarly, one can sum the Taylor expansions above and solve for the second derivative at $x=a$, $f^{\prime\prime}(a)$, from which the numerical approximations of the second partial derivatives are
\begin{equation}
\label{x2ndpartial}
\frac{\partial^{2}h}{\partial x^{2}}=\frac{h_{i+1,j}-2h_{i,j}+h_{i-1,j}}{\Delta x^{2}},%
\end{equation}
\begin{equation}
\label{y2ndpartial}
\frac{\partial^{2}h}{\partial y^{2}}=\frac{h_{i,j+1}-2h_{i,j}+h_{i,j-1}}{\Delta y^{2}}.%
\end{equation}

Let us say for the moment that we are interested in solving the Laplace equation in 2-dimensions
\[
\nabla^{2}h=\frac{\partial^{2}h}{\partial x^{2}}+\frac{\partial^{2}h}{\partial y^{2}}=0,
\]
therefore we replace the second partial derivatives for their Finite Difference numerical equivalents, given by eqs. (\ref{x2ndpartial}) and (\ref{y2ndpartial}), setting $\Delta x=\Delta y=\Delta$ for simplicity, we arrive at
\begin{equation}
h_{i+1,j}+h_{i-1,j}+h_{i,j+1}+h_{i,j-1}-4h_{i,j}=0,
\label{laplace}%
\end{equation}
which is the finite difference formulation of the Laplace equation, coincidentally as will be explained later this very equation corresponds to that of a steady groundwater flow in an homogeneous and isotropic aquifer.

\begin{figure}[h]
\centering
\includegraphics[height=7cm,width=11cm]{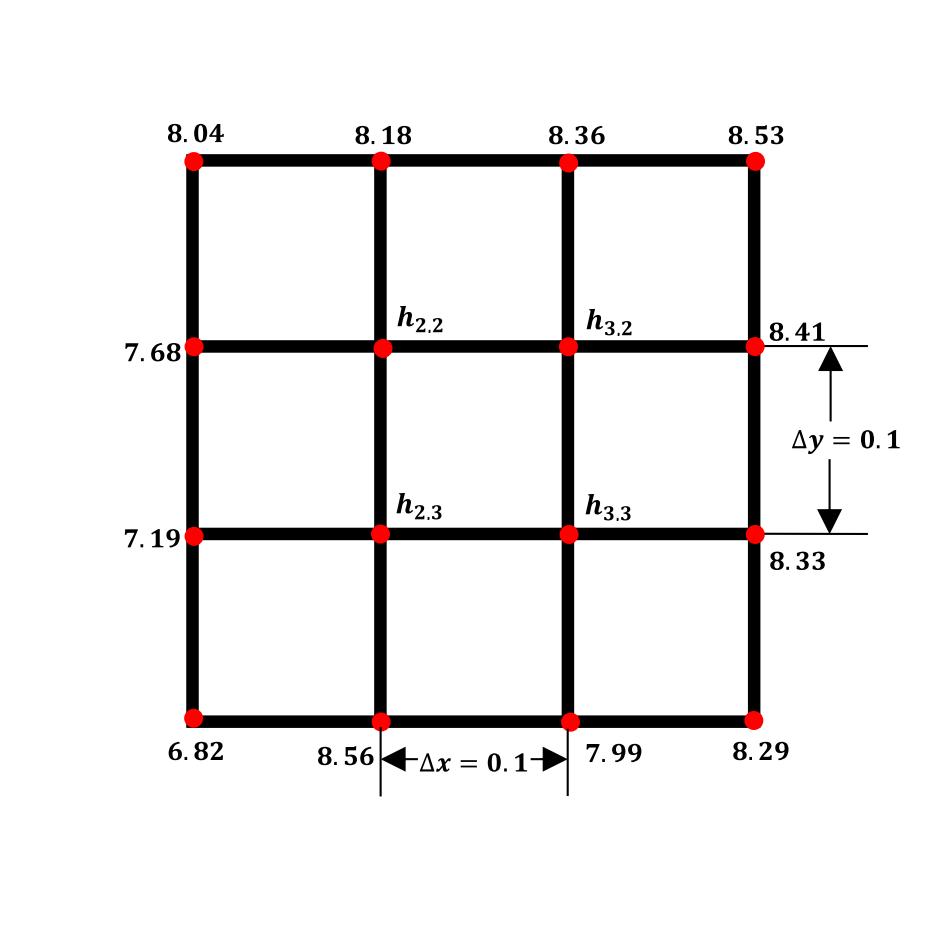}
\caption{{\protect\footnotesize {Finite difference grid for the domain of certain problem.}}}
\label{figinnergrid}
\end{figure}

Along with appropriate boundary values one can compute quantities in the inner part of a grid as exemplified in figure \ref{figinnergrid} in order to solve the problem, in this case we have a set of boundary values on all the external edges of the grid, thus for this case we get
\begin{align*}
h_{1,2}+h_{3,2}+h_{2,1}+h_{2,3}-4h_{2,2}  & =0,\\
h_{1,3}+h_{3,3}+h_{2,2}+h_{2,4}-4h_{2,3}  & =0,\\
h_{2,2}+h_{4,2}+h_{3,1}+h_{3,3}-4h_{3,2}  & =0,\\
h_{2,3}+h_{4,3}+h_{3,2}+h_{3,4}-4h_{3,3}  & =0,
\end{align*}
which is now an algebraic system of equations instead of the original partial differential equation, once we solve this system we have an approximate set of values that describe the behavior of the function $h(x,y)$ over the solution domain.

In practice one make a very large partition of the domain, since it is advisable that the partition size $\Delta$ should be in the interval $(0,1)$ in order to satisfy convergence criteria and to guarantee a good physical approximation, therefore one ends up with a very large system of equations which is very hard if not feasible to handle analitically, so we are in need to implement an iterative numerical scheme of solution. In order to do so we solve eq. (\ref{laplace}) in terms of $h_{i,j}$,
\begin{equation}
\label{5pointoperator}
h_{i,j}=\frac{h_{i-1,j}+h_{i+1,j}+h_{i,j-1}+h_{i,j+1}}{4},
\end{equation}
and with it we compute $h(x,y)$ iteratively all along the grid, thus we obtain ''test'' values from which we construct new ones in every iteration, improving the quality of them on each cycle and tending to get a solution in a prescribed and physically acceptable numerical error tolerance. The quantity in eq. (\ref{5pointoperator}) is often refer to as the five-point operator in the literature, this corresponds to assign at the point of interest the mean value of the four nearest points to it in the nodal array, that is its first neighbors mean value.

There are various iterative methods to solve a system of linear equations, we will refer here to three of them: Jacobi iteration, Gauss-Seidel iteration and Successive Over Relaxation, although we present results only for the second one. These methods in general are the most efficient to solve equations containing a great number of unknowns since they get rid of the need to store data \cite{escobar}.

\subsection{Jacobi Iteration}

We can simply describe the Jacobi procedure as follows, to each interior point in figure \ref{figinnergrid} we apply eq. (\ref{5pointoperator}), this method does not uses a specific order to compute the $h_{i,j}$ values, we use the expresion
\[
h_{i,j}^{m+1}=\frac{h_{i-1,j}^{m}+h_{i+1,j}^{m}+h_{i,j-1}^{m}+h_{i,j+1}^{m}}{4},
\]
where $m$ labels the previous values to the iteration that it is been computed and labeled by $m+1$. First one needs to propose a set of initial values, generally by guess but physically based on the boundary values of the problem. Following the example on figure \ref{figinnergrid} one is led to propose
\[
h_{2,2}^{0}=8,\text{ }h_{3,2}^{0}=8.5,\text{ }h_{2,3}^{0}=7\text{ and } h_{3,3}^{0}=8,
\]
as an initial solution from which after three iterations we get
\[
h_{2,2}^{3}=7.84,\text{ }h_{3,2}^{3}=8.19,\text{ }h_{2,3}^{3}=7.69\text{ and } h_{3,3}^{3}=7.96,
\]
one continues with the iterations until some physically based prescribed error tolerance is met.

\subsection{Gauss-Seidel Iteration}

Unlike the method described above Gauss-Seidel iteration is performed in an orderly fashion, we compute across the grid from left to right and downwards line per line so that on each node we can used the newest values to compute the next one during the same iteration, in this way we accelerate the convergence of the computation. The iterative expression one uses is
\[
h_{i,j}^{m+1}=\frac{h_{i-1,j}^{m+1}+h_{i,j-1}^{m+1}+h_{i+1,j-1}^{m}+h_{i,j+1}^{m}}{4},
\]
the movement with which we traverse the grid is totally similar to that of reading a book's page. The fact we use the new available values makes possible that this method has better performance than the simple Jacobi iteration \cite{grossmann}, as will be discussed in section \ref{sec_numerical}.

\subsection{Successive Over Relaxation}

To implement the SOR method we must define the residue, $c$, this quantity is given by the change between two successive iterations in the Gauss-Seidel method, thus
\begin{equation}
c=h_{i,j}^{m+1}-h_{i,j}^{m},
\label{residue}
\end{equation}
in every successive Gauss-Seidel iteration the residue softly vanishes on each node with the use of the new computed values, therefore converging more and more rapidly. In this method the residue is multiply by a relaxation factor $\omega$ with $\omega \geq 1$ such that the newly computed value is given by $h_{i,j}^{m+1}=h_{i,j}^{m}+\omega c$, in practice this means that in the SOR formula we substitute the definition of the residue along with the Gauss-Seidel expression for $h_{i,j}^{m+1}$ which results in
\[
h_{i,j}^{m+1}=\left(1-\omega \right) h_{i,j}^{m}+ \omega \frac{h_{i-1,j}^{m+1}+h_{i,j-1}^{m+1}+h_{i+1,j-1}^{m}+h_{i,j+1}^{m}}{4}.
\]

We say then that one works in the over relaxation scheme since we add more residue to the value at hand, if on the contrary we have $0 \leq \omega \leq 1$ the updated value of $h_{i,j}^{m+1}$ is said to be under relaxed and it corresponds to a weighted average of the values used to compute the new one. Under relaxation is generally employed in order to make a non-convergent system converge by damping the oscillations in the computed values \cite{grossmann}.

\section{Groundwater Flow Model}
\label{sec_model}

In order to obtain the mathematical model of groundwater dynamics, we assume that water, or any other fluid under study, obeys Darcy's law of flow through porous media, that is to say that the fluid is incompressible, that its Reynolds number $N_{R}<1$, experimentally it has been determined that any fluid with a Reynolds number less than 1 satisfies Darcy's law and also that it is a good approximation for fluids with $1\leq N_{R} < 10$, and additionally one requires the flow to be laminar. In its simple and modern form Darcy's law relates $\mathbf{q}$ the specific discharge vector [m/s], $\mathbf{\sigma}$ the hydraulic conductivity tensor [m/s] and $h(\mathbf{r})$ a scalar field standing for the total head [m] by means of
\begin{equation}
\label{darcy}
\mathbf{q}=- \mathbf{\sigma}\cdot\nabla h,
\end{equation}
where the negative sign indicates that the flow of water is in the direction of decreasing head.

Assuming that our coordinate axes coincide with the principal axes of the conductivity tensor then for an isotropic homogeneous medium $\mathbf{\sigma}$ has a diagonal form with three identical constant elements and Darcy's law simplifies to $\mathbf{q}=-\sigma\nabla h$, nevertheless the more realistic configuration involves an inhomogeneous and anisotropic hydraulic conductivity, in this cases this tensor can be represented as a $3 \times 3$ diagonal matrix
\[
\mathbf{\sigma}=\left(
\begin{array}
[c]{ccc}%
\sigma_{x}(x,y,z) & 0 & 0\\
0 & \sigma_{y}(x,y,z) & 0\\
0 & 0 & \sigma_{z}(x,y,z)
\end{array}
\right),
\]
where each component has a local value depending on the position.

The equation governing groundwater flow can be obtained by the Control Volume approach \cite{keith} and it states that $\nabla \cdot \mathbf{q}=S_{s}\frac{\partial h}{\partial t}+ W$, where $S_{s}$ is the specific storage and $W$ represents flow in or out of the control volume, these two terms are usually written in a single function $f(\mathbf{r},t)$ called infiltration and represents the presence of sources or wells, using eq. (\ref{darcy}) this is written as
\begin{equation}
\label{floweq}
\frac{\partial}{\partial x}\left(  \sigma_{x}\frac{\partial h}{\partial x}\right)  +\frac{\partial}{\partial y}\left(  \sigma_{y}\frac{\partial h
}{\partial y}\right)  +\frac{\partial}{\partial z}\left(  \sigma_{z}%
\frac{\partial h}{\partial z}\right)  =-f\left(  \mathbf{r},t\right),
\end{equation}
which represents groundwater flow in an inhomogeneous anisotropic aquifer, this equation is solved in order to obtain the potential head scalar field $h$, from which we determine the rates and directions of flux from the specific discharge vector. In the case of steady flow in a homogeneous isotropic and confined aquifer, the continuity or conservation law assures that eq. (\ref{floweq}) reduces to the Laplace equation.

\begin{figure}[h]
\centering
\includegraphics[height=7cm,width=11cm]{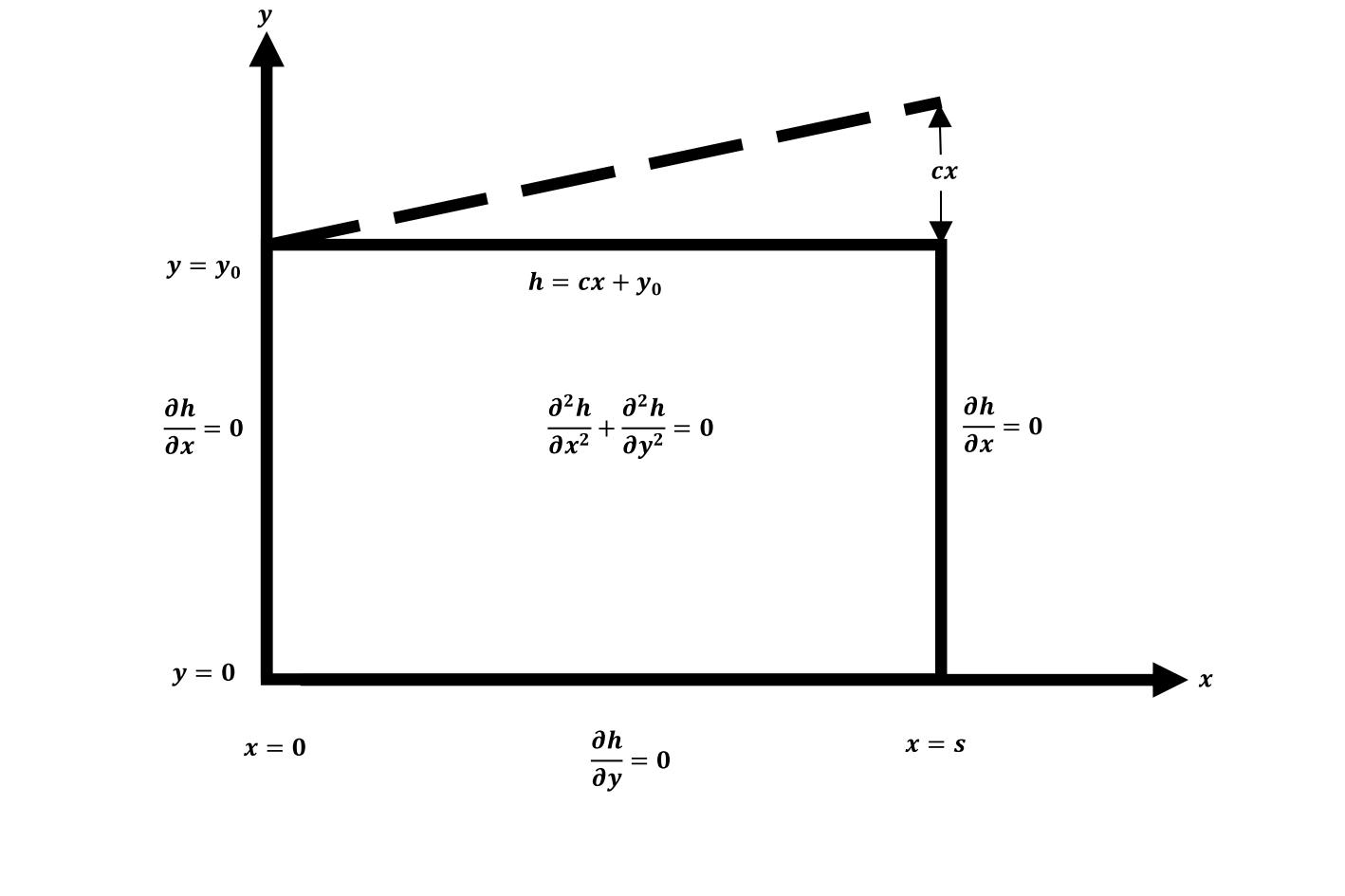}
\caption{{\protect\footnotesize {Sketch of the mathematical model and boundary conditions for the regional groundwater system studied by Toth.}}}
\label{figboundaries}
\end{figure}

\begin{figure}[h]
\centering
\includegraphics[height=7cm,width=11cm]{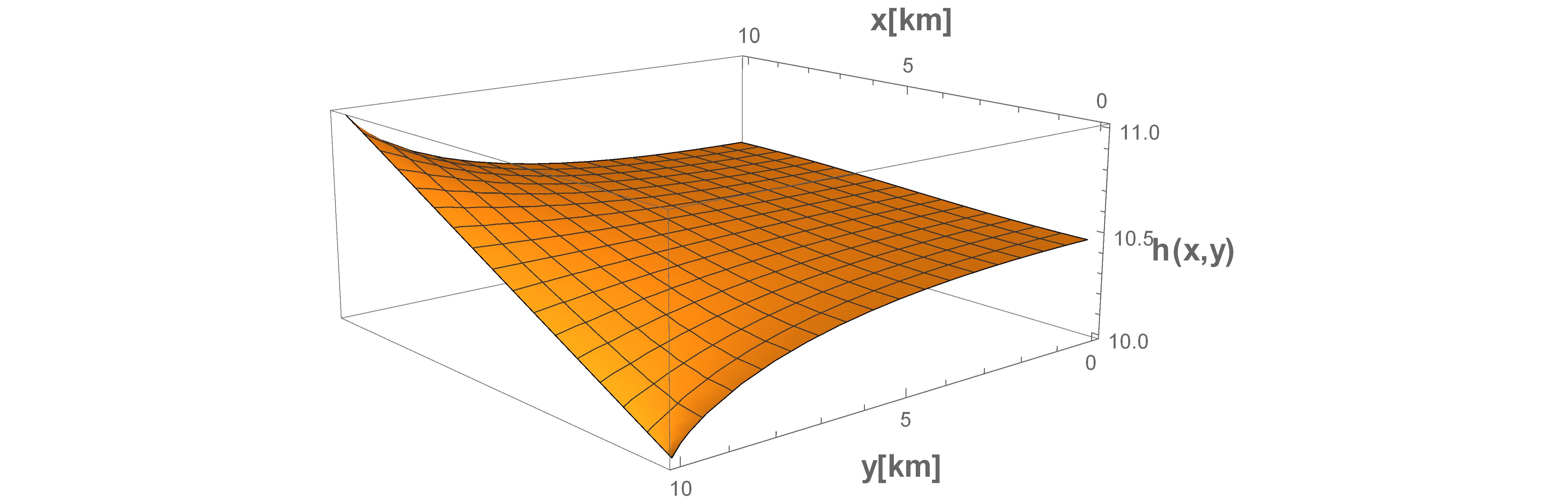}
\caption{{\protect\footnotesize {Profile of Toth's analytical solution for the aquifer describe above and the domain configuration depicted in fig. \ref{figboundaries} with $y_{0} = 10$ km, $s = 10$ km and  $c = 0.1$.}}}
\label{figtoth}
\end{figure}

In the following we restrict ourselves to the case of a 2-dimensional confined aquifer and to the geometry proposed by Toth \cite{toth1,toth2}, where he studies a confined aquifer formed of a homogeneous, isotropic, porous medium with an underlaying layer of impermeable rock and a water table of linear slope. The system consists of a small watershed bounded on one side by a topographic high and on the other side by a mayor stream, both boundaries mark regional groundwater divides and they are treated mathematically as no-flow boundaries as well as the lower boundary, the upper boundary of the mathematical model is the horizontal line at height $y=y_{0}$ although the water table of the system lies above this line and the head along this boundary is assume to variate linearly as $h(x,y_{0})=cx+y_{0}$, where $c$ is the slope of the water table. Thus, the domain of the problem is an approximation to the actual shape of a saturated flow region, in figure \ref{figboundaries} we depict the cross section of the watershed along with the boundary conditions. The solution for this system in the simplified setup described above is given by
\begin{equation}
h(x,y)=y_{0}+\frac{cs}{2}-\frac{4cs}{\pi^{2}}\sum_{m=0}^{\infty}\frac
{\cos\left[  \left(  2m+1\right)  \frac{\pi x}{s}\right]  \cosh\left[  \left(
2m+1\right)  \frac{\pi y}{s}\right]  }{\left(  2m+1\right)  ^{2}\cosh\left[
\left(  2m+1\right)  \frac{\pi y_{0}}{s}\right]  },
\label{soltoth}
\end{equation}
needles to say that the homogeneity and isotropy simplifications are not compulsory to obtain a numerical solution. In figure \ref{figtoth} we depict the profile of Toth's analytical solution.

It is in order to say that the study of analytical solutions for steady and transient flow, though maybe non totally realistic, can give us a great deal of insight of an aquifer and groundwater flow systems. For instance the stationary solution can be used to determine the response time of a system under transient flow, one of the methods to compute this time consists in solving both equations, for steady and transient flow, with which the response time will be the amount of time required for the transient solution to approach to the stationary one within a predefined tolerance \cite{simpson1,simpson2,rousseau}. Measure of this time scale is of critical importance since it can help us to define whether or not it is appropriate to use a transient flow model, if the time scale of interest is greater than the time response of the system then it is adequate to use a steady-state mathematical model and inversely \cite{haitjema}. Field work does not allow much room for error when establishing the computational model, in most cases is too expensive to redo the model should second guesses or doubts appear, at this point some simple analytical solutions, like those for one dimensional or radial flow, can provide us with a clear idea of the fundamental behavior of more complex or real world groundwater flow systems, thus serving as a tool both to decide between a steady or a transient model and  to fine tune the parameters \cite{haitjema}.

Moreover problems susceptible of being analytically solved are still the battleground to probe new predictive techniques, as the mean action integral recently proposed in \cite{jazaei}. Also, in the practice we are presented with the problem of sustainable exploitation, thus one is interested in the stationary solution and the response time of the system to the end of predicting the time it would take an aquifer to reach equilibrium once pumping has been introduced, specially if we want the pumping to continue essentially in an indefinite form, this time window in much cases ranges the millennia scale \cite{bredehoeft}.


\section{Numerical Implementation}
\label{sec_numerical}

Making use of the results of section \ref{sec_iterativeformulation} we will rewrite the general flow equation for a 2-dimensional aquifer as a purely algebraic discrete expression as follows, let us consider first the general case of an inhomogeneous anisotropic confined aquifer in steady flow, thus the right hand side of eq. (\ref{floweq}) vanishes, also for the moment we leave the components of the hydraulic conductivity tensor unfixed, upon using eqs. (\ref{xpartial})-(\ref{y2ndpartial}) and solving for $h_{i,j}$ we get
\begin{equation}%
\begin{array}
[c]{c}%
h_{i,j}=\frac{1}{2\left(  \sigma_{x}+\sigma_{y}\right)  }\left[  \left(\sigma_{x}+\frac{\Delta}{2}\partial_{x}\sigma_{x}\right)  h_{i+1,j}+\left(\sigma_{x}-\frac{\Delta}{2}\partial_{x}\sigma_{x}\right)  h_{i-1,j}+\left(\sigma_{y}+\frac{\Delta}{2}\partial_{y}\sigma_{y}\right)  h_{i,j+1}+\left(\sigma_{y}-\frac{\Delta}{2}\partial_{y}\sigma_{y}\right)  h_{i,j-1}\right],
\end{array}
\label{floweqdiscrete}%
\end{equation}
where we have set $\Delta x=\Delta y=\Delta$ for simplicity.

To compute the solution of this system we use the Gauss-Seidel iteration method since it will converge faster than the simple Jacobi iteration, also given the typical values of $h$ and the components of the $\sigma$ tensor the algebraic equations represented by (\ref{floweqdiscrete}) are well posed and behaved, making unnecessary to use the Under Relaxation method, on the other hand, also in terms of the rate of convergence, according to these facts, one does not gain much if the Over Relaxation method were to be used.

We implement the numerical solution using Python 3.6, recalling we use a second order approximation let us say that we take $\Delta=0.1$, therefore our approximation gives us a numerical result accurate to at most $\mathcal{O}(2) \sim10^{-2}$ which means according to \cite{scarborough} that our calculations have three significant figures of accuracy, therefore we are forced to implement the iterative scheme within a loop that gradually rises the level of accuracy of the results until a prespecified error tolerance, $e_{s}$, is achieved and with this the desired number of significant figures is reached.

\begin{figure}[h]
\centering
\includegraphics[height=7cm,width=11cm]{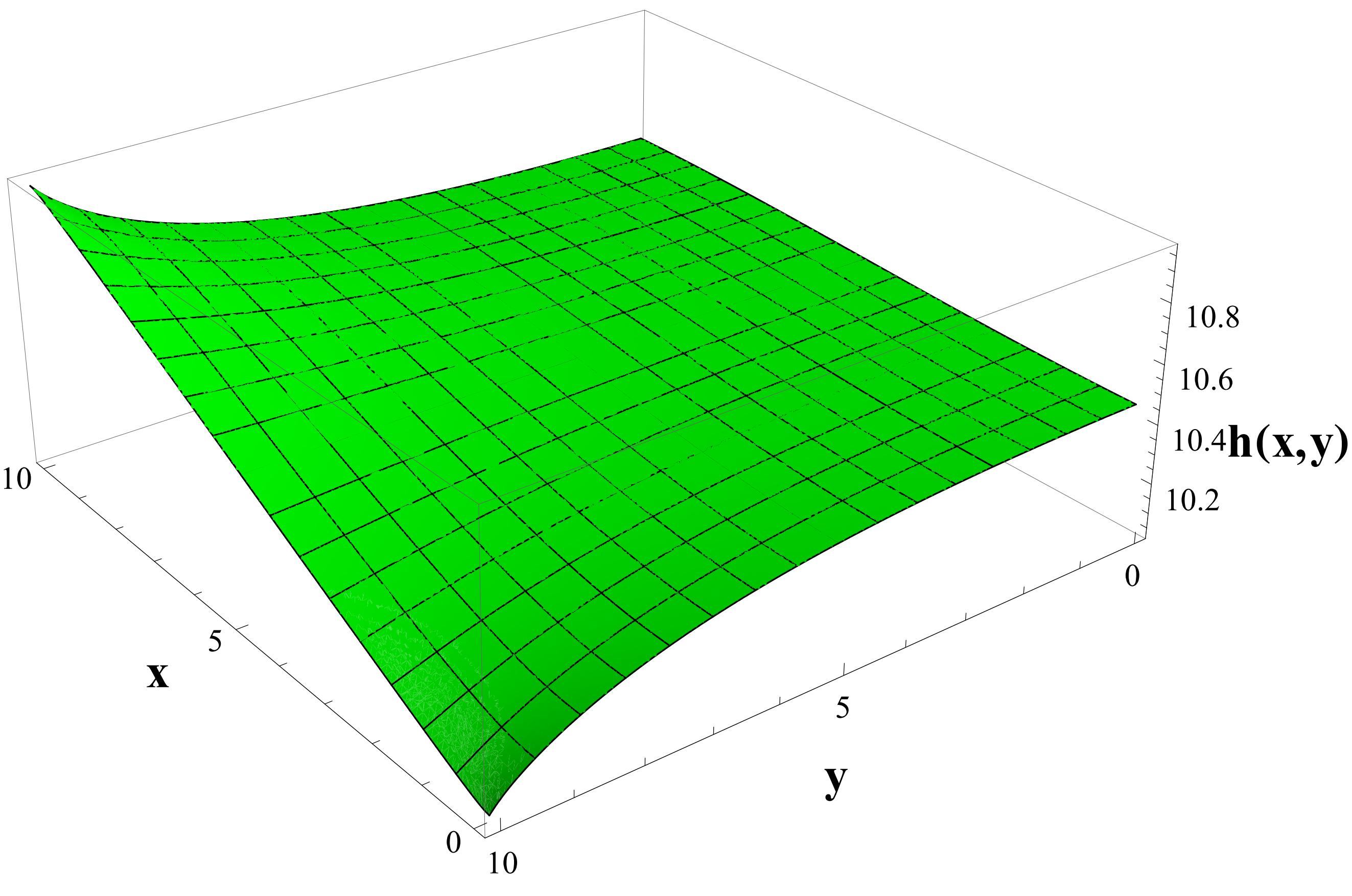}
\caption{{\protect\footnotesize {Profile of the numerical solution to the isotropic aquifer problem sketch by Toth domain configuration depicted in fig. \ref{figboundaries} with $y_{0} = 10$ km, $s = 10$ km and  $c = 0.1$, $\Delta = 0.1$ km and $e_{s}=10^{-8}$ .}}}
\label{figtothnum}
\end{figure}

\begin{figure}[h]
\centering
\includegraphics[height=7cm,width=11cm]{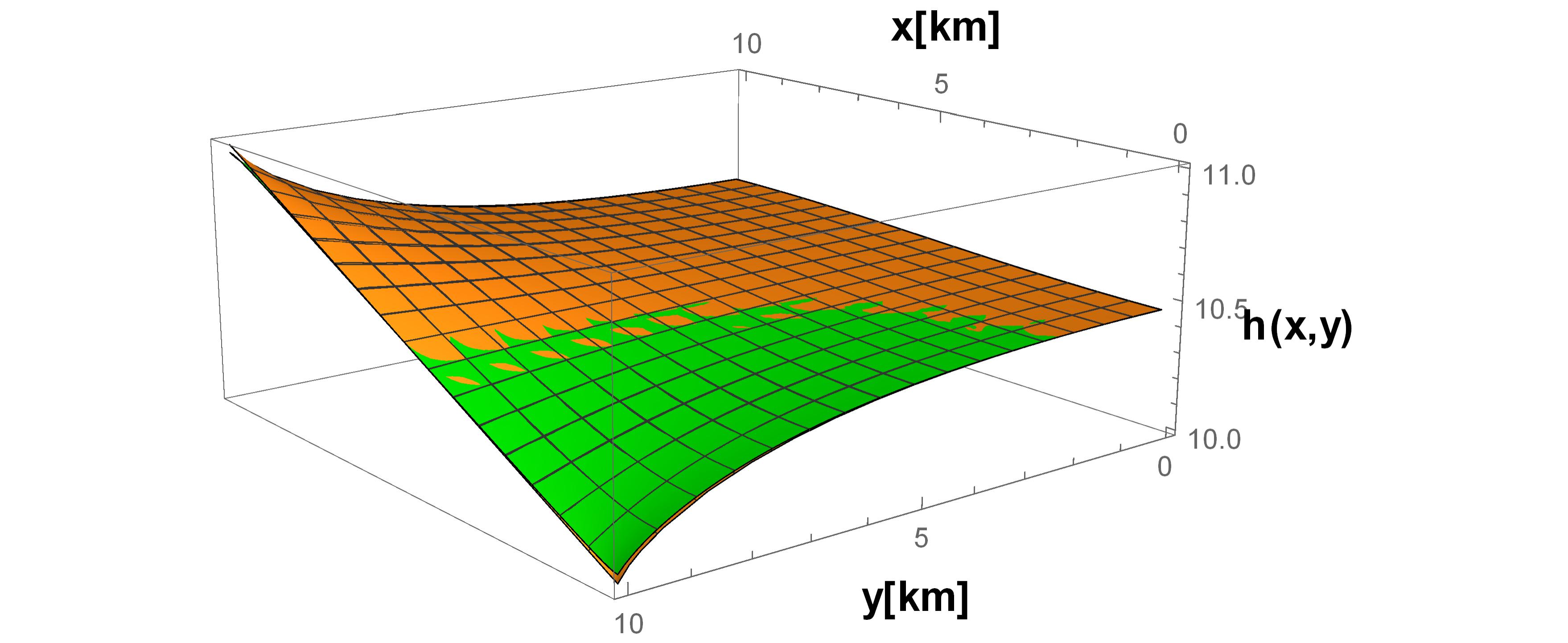}
\caption{{\protect\footnotesize {Overlapping of analytical and numerical solution profiles, here we have taken $y_{0} = 10$ km, $s = 10$ km and  $c = 0.1$, $\Delta = 0.1$ and $e_{s}=10^{-8}$ .}}}
\label{figsuperpos}
\end{figure}

For the aquifer setup described at the end of the previous section we have $\sigma =constant$, thus eq. (\ref{floweqdiscrete}) reduces to
\[
h_{i,j}=\frac{1}{4}\left[  h_{i-1,j}+h_{i+1,j}+h_{i,j-1}+h_{i,j+1}\right],%
\]
which is nothing else than the mean value of the four neighboring points of $h_{i,j}$. Solving these numerically, we get the profile shown in figure \ref{figtothnum}, where we depict the head scalar field we obtain with our code, also figure \ref{figsuperpos} shows the super position of the analytical and numerical profiles, as we can see they are in great agreement showing overlapping in most of the domain region except in the far edges, where this behavior is to be expected given the numerical approximation scheme and its unavoidable natural error.

\begin{figure}[h]
\centering
\includegraphics[height=7cm,width=11cm]{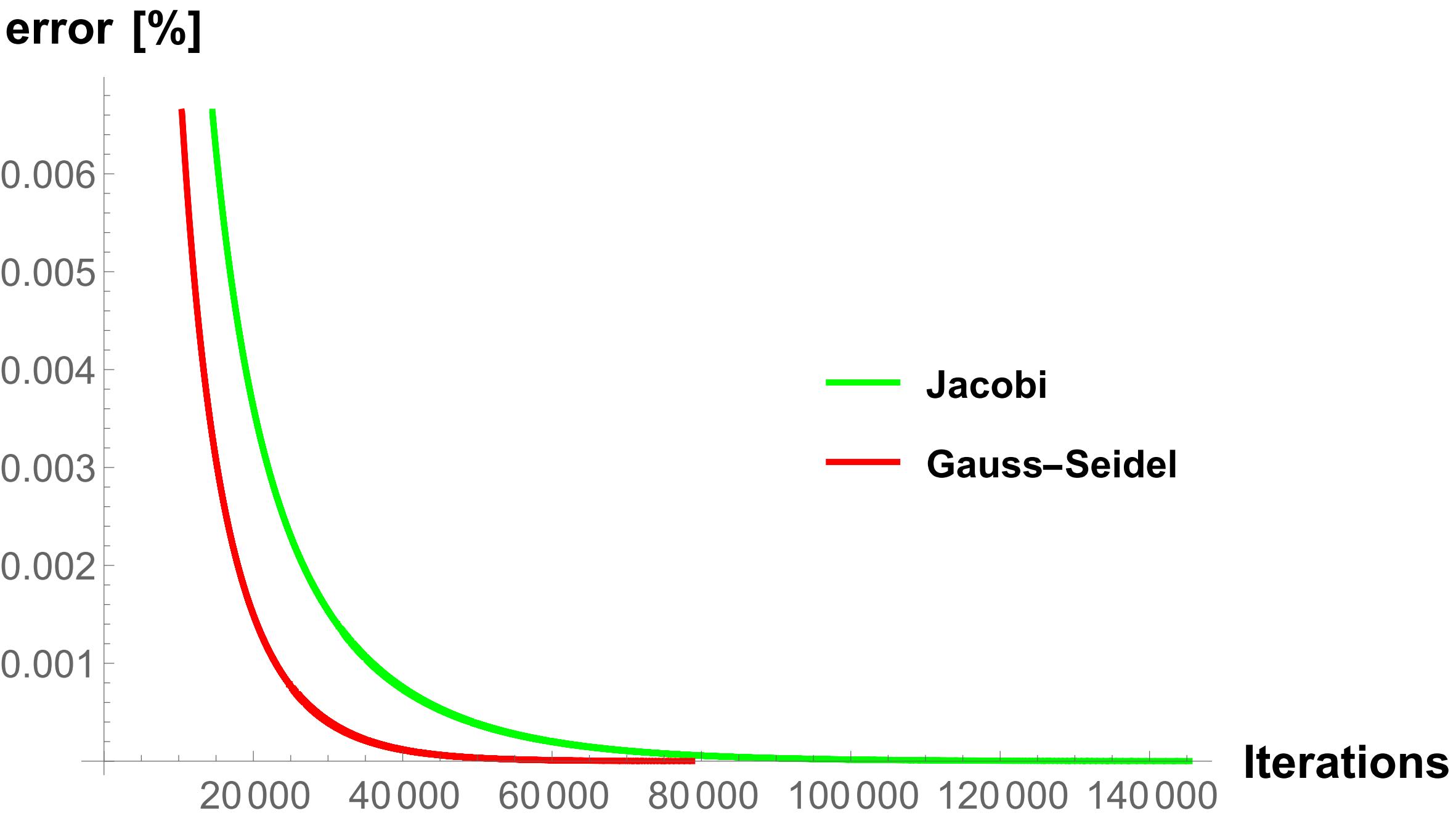}
\caption{{\protect\footnotesize {Convergence curves of the Jacobi and Gauss-Seidel numerical solution methods for the case of flow in an isotropic homogeneous aquifer, we have taken $y_{0} = 10$ km, $s = 10$ km and  $c = 0.1$, $\Delta = 0.1$ and $e_{s}=10^{-8}$ .}}}
\label{figerror}
\end{figure}

In figure \ref{figerror} we show the convergence curve of both Jacobi and Gauss-Seidel numerical solution methods for this homogeneous isotropic aquifer system, where we can see that Gauss-Seidel method reaches the prespecified error tolerance $e_{s}$ at approximately half the number of iterations with respect to the Jacobi method. This is no surprise if we recall the definition of the asymptotic rate of convergence of an iterative procedure, given by
\[
\lambda = - \ln \rho\left(\mathbf{B}\right),
\]
where $\mathbf{B}$ is the iteration matrix of the procedure and $\rho$ is its spectral radius, then since $\rho\left(\mathbf{B}_{GS}\right)=\rho\left(\mathbf{B}_{J}\right)^{2}$ for these two methods \cite{quarteroni}, we have as a general result that $\lambda_{GS}/\lambda_{J}=2$. This implies that in general Gauss-Seidel procedure converges twice as fast as Jacobi procedure, therefore in order to reach certain error tolerance it requires half the iterations with respect to the later.

\begin{figure}[h]
\centering
\includegraphics[height=7cm,width=11cm]{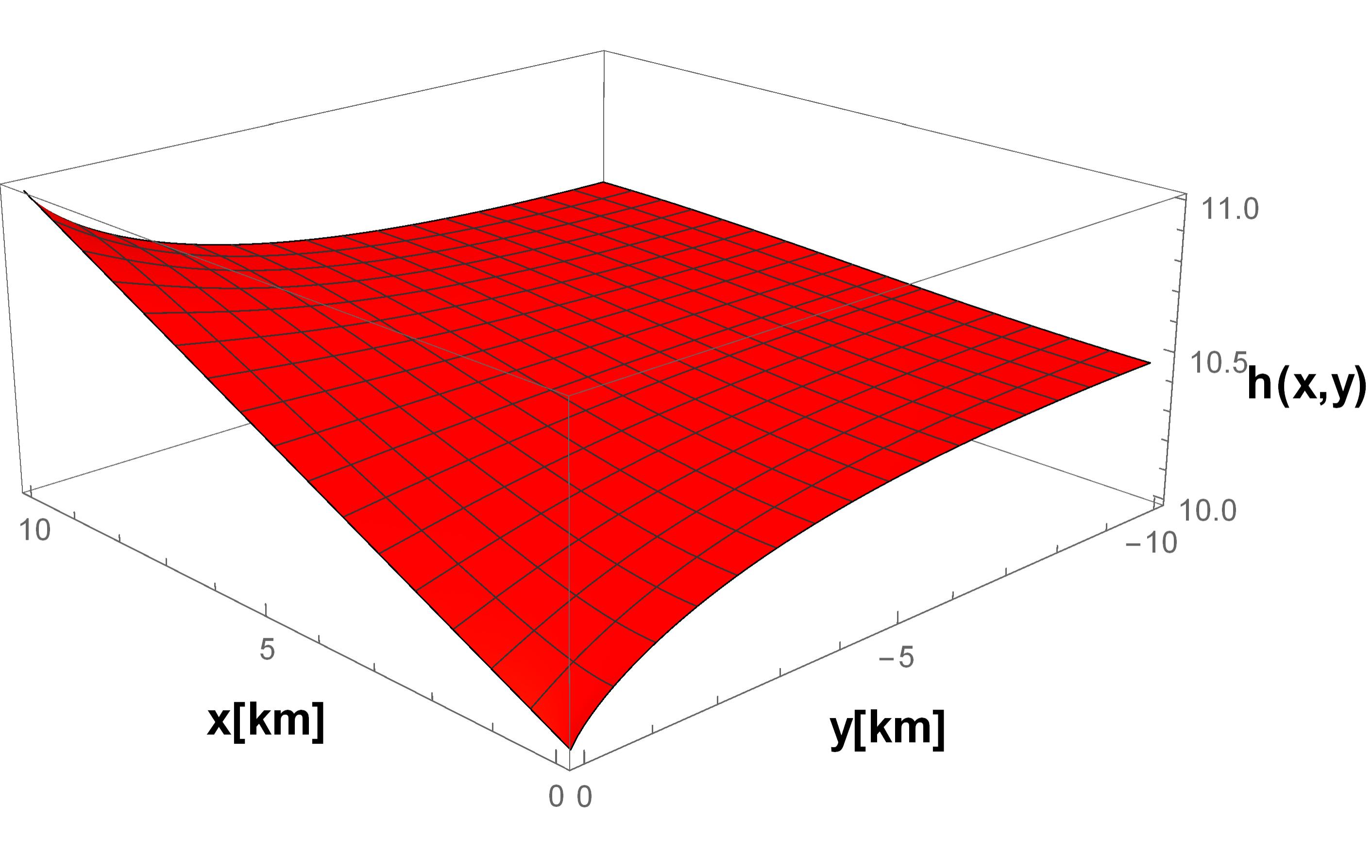}
\caption{{\protect\footnotesize {Profile of the numerical solution to the homogeneous aquifer problem sketch by Toth domain configuration depicted in fig. \ref{figboundaries} with $y_{0} = 10$ km, $s = 10$ km and  $c = 0.1$, $\Delta = 0.1$ km and $e_{s}=10^{-8}$ .}}}
\label{fighomoanis}
\end{figure}

\begin{figure}[h]
\centering
\includegraphics[height=7cm,width=11cm]{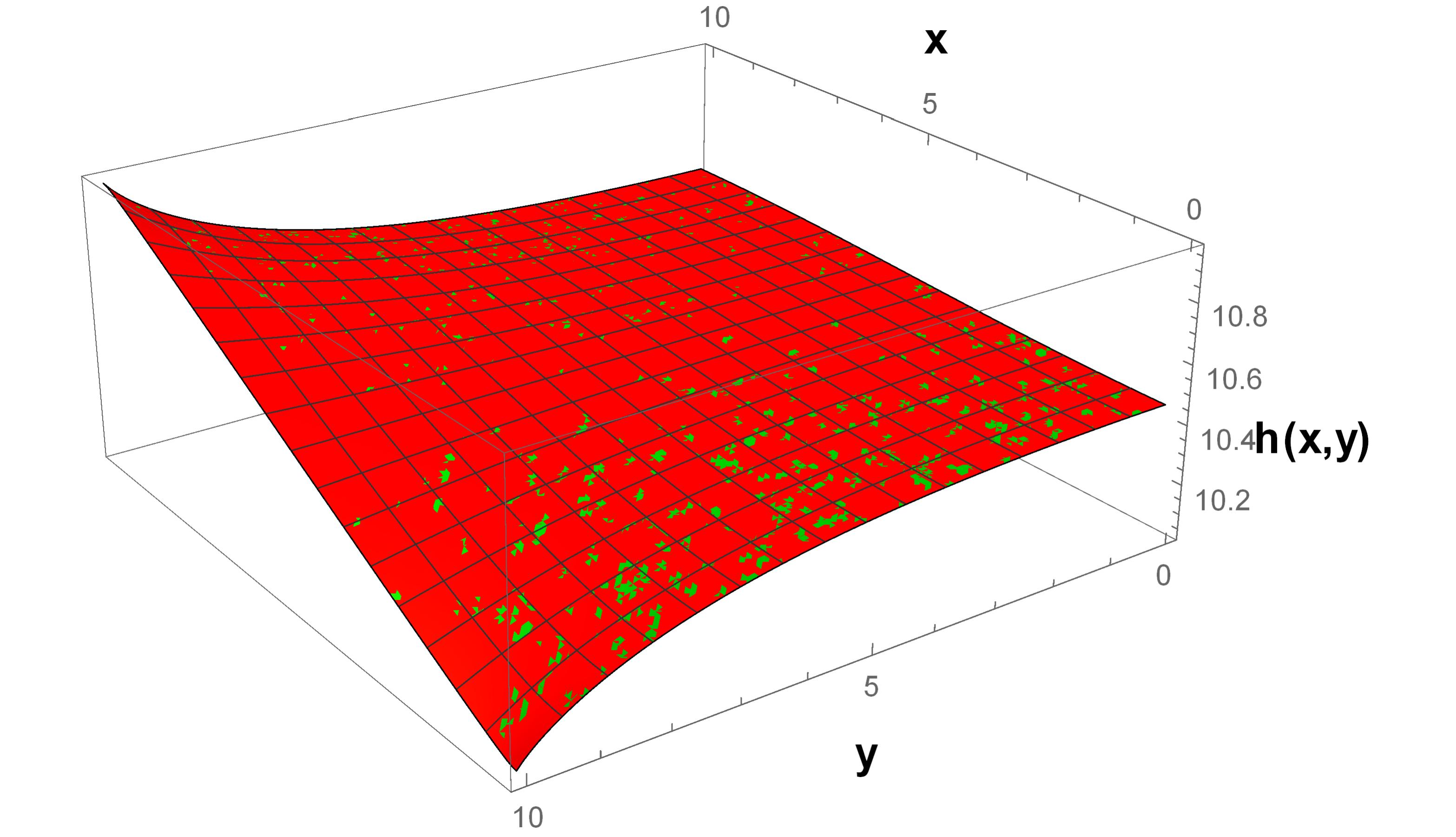}
\caption{{\protect\footnotesize {Overlapping of the numerical solutions for the isotropic and homogeneous anisotropic cases, here we have taken $y_{0} = 10$ km, $s = 10$ km and  $c = 0.1$, $\Delta = 0.1$ km and $e_{s}=10^{-8}$ .}}}
\label{figsuperposhomoanis}
\end{figure}

These results make us confident to apply our computational code to a 2D-cross section of an homogeneous anisotropic aquifer within the same approximations as the problem described above. We take for instance the Ayamonte-Huelva aquifer, localized in the Spanish province of Huelva in Andalucía County, whose hydrogeological characteristics suggest the presence of materials such as  sand, sandstone, clay, slate, gravel, lime and chalk \cite{atlas,gonzalez}. As stated before the hydraulic conductivities of the soil and rocks play a main role in the dynamics of the system since conductivity depends on physical factors as the porosity, size and form of the grain, as well as the geometrical distribution of the particles in the media and the physical properties of the fluid, generally clay materials exhibit low values of hydraulic conductivity while sandstone and gravel have high values. The conductivity of saturated regions can be determined by various techniques including mathematical calculations, laboratory methods, tracer tests, well tests, auger hole tests and pumping tests of wells. Here we will used the representative values of vertical and horizontal hydraulic conductivity, $\sigma_{y}$ and $\sigma_{x}$ in our convention, for the rock types present in the Ayamonte-Huelva aquifer \cite{domenico}, they are $0.003721755$ m/s and $0.003721761$ m/s respectively, we computed these values in a rough approximation as the weighted average of the conductivity of each material composing the soil in the aquifer mentioned before. Under these homogeneity and anisotropy conditions eq. (\ref{floweqdiscrete}) is written as
\[
h_{i,j}=\frac{1}{2\left(  \sigma_{x}+\sigma_{y}\right)  }\left[  \sigma_{x}h_{i-1,j}+\sigma_{x}h_{i+1,j}+\sigma_{y}h_{i,j-1}+\sigma_{y}h_{i,j+1}\right],
\]
which in resemblance to the expression for the isotropic case this last equation amounts to a weighted mean of the first neighbors of $h_{i,j}$, in figures \ref{fighomoanis} and \ref{figsuperposhomoanis} we depict the profiles of the numerical solution for this case, within the same boundary conditions as before, as well as its superposition with the isotropic case fig. \ref{figtothnum}.

\begin{figure}[h]
\centering
\includegraphics[height=7cm,width=11cm]{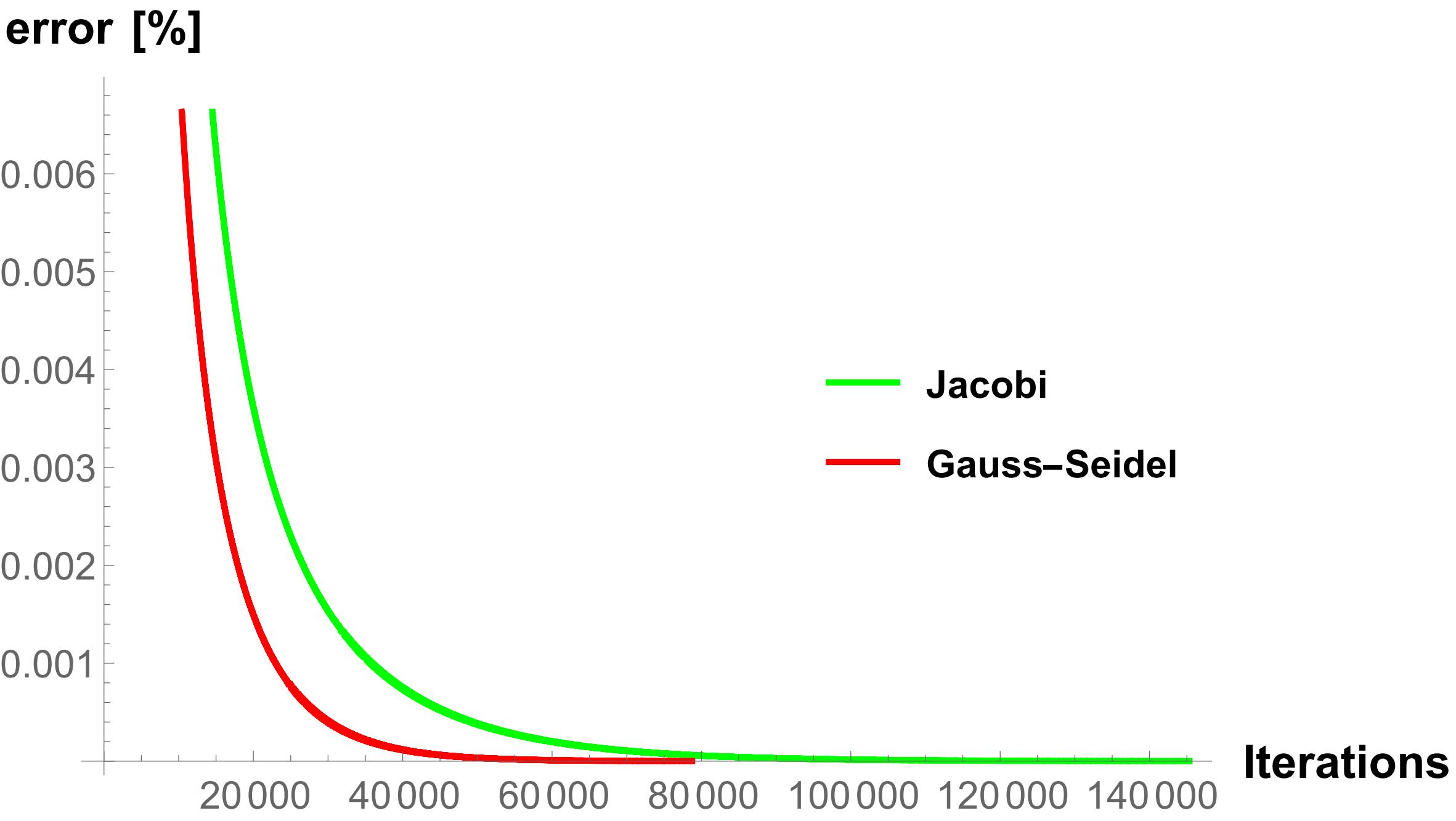}
\caption{{\protect\footnotesize {Convergence curves of the Jacobi and Gauss-Seidel numerical solution methods for the case of flow in an homogeneous anisotropic aquifer, we have taken $y_{0} = 10$ km, $s = 10$ km and  $c = 0.1$, $\Delta = 0.1$ and $e_{s}=10^{-8}$ .}}}
\label{figerrorhomoanis}
\end{figure}

As we can see from figures \ref{fighomoanis} and \ref{figsuperposhomoanis} heads for both cases, isotropic and anisotropic aquifer, are basically equivalent which is to be expected since the numerical values of the horizontal and vertical conductivities we computed are almost equal, also their significant figures are the same for both values in the numerical model due to the error tolerance we prescribed; additionally the mean deviation of both calculations is order $10^{-9}$, thus lies bellow the accuracy we work with. This tells us that in practice we can treat this aquifer as an isotropic homogeneous one within our approximation. In figure \ref{figerrorhomoanis} we also show the convergence curves upon implementation of the Gauss-Seidel and Jacobi iterative numerical procedures.

\begin{figure}[h]
\centering
\includegraphics[height=11cm,width=11cm]{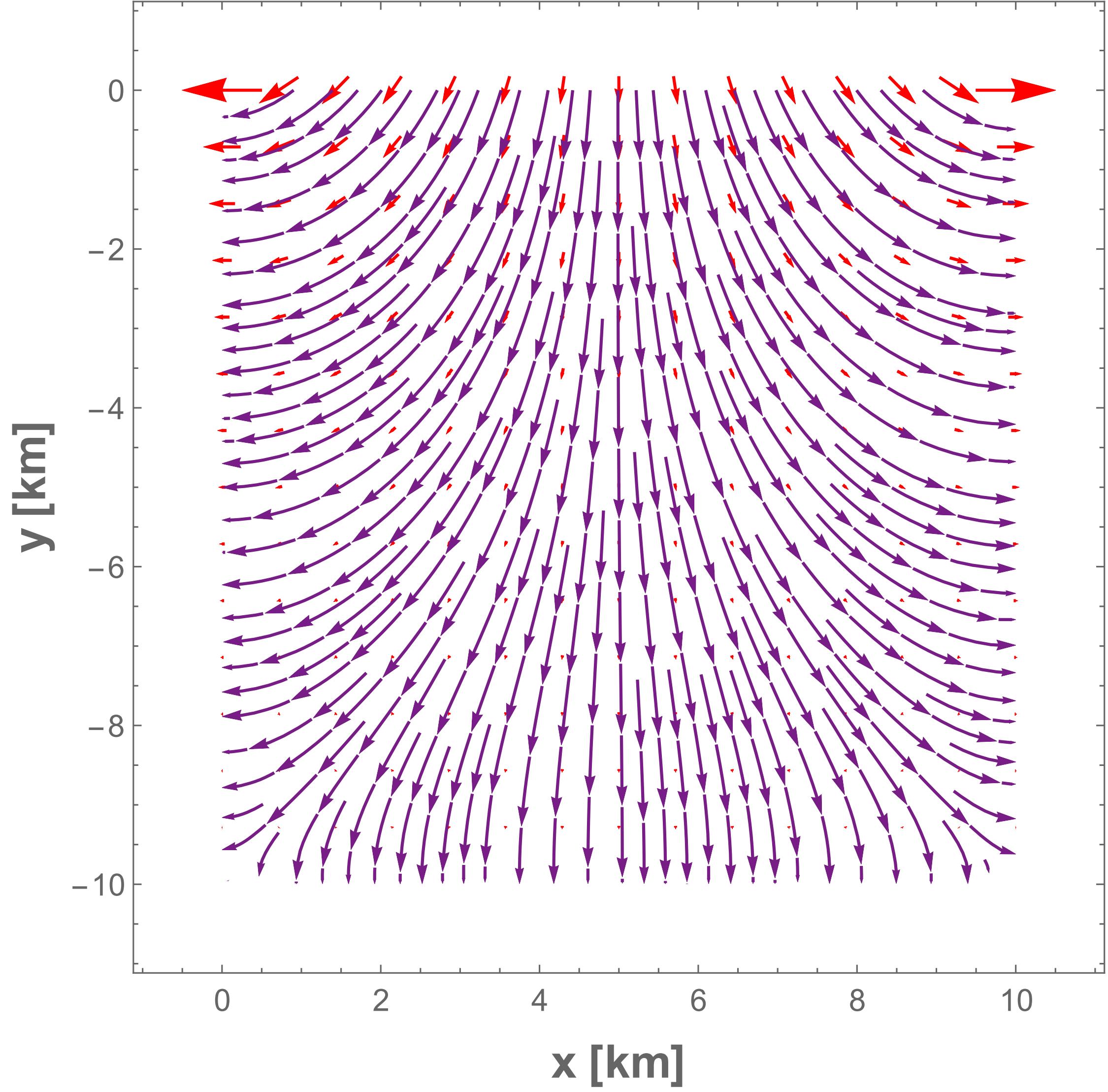}
\caption{{\protect\footnotesize {Directional field (red) of the water flow inside the aquifer according to eq. (\ref{darcy}), also we depict the stream lines (purple) of the flux, for the isotropic homogeneous case with $y_{0} = 10$ km, $s = 10$ km and  $c = 0.1$, $\Delta = 0.1$ km and $e_{s}=10^{-8}$ .}}}
\label{figdirectionalfield}
\end{figure}

From fig. \ref{figtothnum} we can appreciate the qualitative behavior of the flow, recalling from Darcy's law that the negative sign in eq. (\ref{darcy}) means that the groundwater flows in the direction of head loss, also the exact stream lines and the specific discharge vector can be seen from fig. \ref{figdirectionalfield}. It is easy to see that  at the boundary $x=0$  km water flows to the left from the water table and in an analog way along the line $x=10$ km water goes right from the water table, towards the side boundaries of the aquifer. Along the boundary $y=0$ km flow is basically downwards with a slightly sideway deviation to the vertical boundaries of the aquifer. The $y=-10$ km receives a great amount of water, since as can be seen from the directional field approximately a third of the stream lines ends in this boundary. Using the numerical results we compute the average discharge vector, which in the units we used represents also the velocity of the flux, resulting in $\mathbf{q}= \left[ 0.001406 \mathbf{\hat{x}} - 0.010008\mathbf{\hat{y}} \right]$ m/s, this tells us, with in this approximation, that the most adequate region the exploitation of this aquifer is along the boundary on the righthand side, also upon computation of the root mean square of $\mathbf{q}$ we know that in general water flows at $v=0.019362$ m/s, this figure is important since it helps us to estimate in an approximate fashion the volume of water we can extract per unit time.

We should state that while groundwater flow in aquifers can be modeled via specialized softwares, based on finite differences and directly oriented to the problem, as Visual MODFLOW and MODFLOW-SURFACT in our opinion they present certain limitations, for instance MODFLOW requires additional subroutines, not included as part of the suit, in order to discretize the data one feeds to the software (via text archives) in the form it needs; on the other hand Visual MODFLOW's major disadvantage resides in the numerical formulation, it is impossible to "interpolate" between cell values when there is no measured or fed value available, this software drys out the cell assigning to it a zero head value, therefore producing a computational error; another problem is that its numerical resolution allows only to work with regular aquifers, generally rectangular or prismatic, therefore to model irregular, anisotropic or heterogeneous aquifers via this software is very difficult. All this situations can be overcome with careful, though laborious, coding and detailed calibration of the computer model, as well as a sufficiently accurate and precise finite difference scheme \cite{quarteroni}.

Also we must emphasize the role computer groundwater models have gained in a wide variety of applications from critic environmental problems to direct industrial applications, consider for instance a contaminated aquifer by heavy metals for which having a 3-dimensional transport model in the case of heterogeneous flow is of vital importance in order to implement soil remediation strategies, these models are generally described in terms of the water balance equation \cite{hansen}, a totally analogous treatment to the flow equation derived from Darcy's law. In the same sense an adequate groundwater model is of the uppermost relevance in productive activity like the extraction of rocks from a quarry. Understanding of the interaction between a quarry and groundwater is fundamental since rock operators in general excavate as deeper as the deposit allows, given this situation the water table of the formation is usually reached and pumping systems are implemented in order to dewater the aquifer, thus depressing the phreatic level beneath the bottom of the pit. Currently there are works such as \cite{barthelemy} where interaction between quarry and groundwater is predicted by means of mathematical indexes developed from a multifold perspective, the key idea of using an index is that it quantifies in a single number the information of a 3-dimensional computational model could provide us with.

\section{Conclusions}
\label{conclusions}

We have studied the numerical solution of the steady groundwater flow equation obtained from application of Darcy's law for flux in porous media, the precise computation of the discrete version of such equation was given considering the general case of an inhomogeneous anisotropic confined aquifer with steady flow. We tested our second order approximation scheme by confronting numerical results with a known analytical solution. Application of this methodology in the case of the Ayamonte-Huelva aquifer showed that and its actual hydraulic parameters, within the second order finite difference approximation of the differential operators and with the tolerance we work with, this aquifer can be regarded as comprised of an homogenous and isotropic medium since consideration of anisotropy produces negligible changes in the numerical values of the hydraulic potential (with respect to the isotropic case) and thus in the generic behaviour of water flow. We should emphasized that this is valid for the vertical and horizontal hydraulic conductivity values, computed as the weighted mean of the conductivities reported in the literature for the components of the aquifer's soil. With this setup we were able to determine a first approximation of the optimal location for exploitation of the aquifer as well as an estimation of the flux. Further work is required, we are preparing a 3-dimensional fourth order code to model Ayamonte-Huelva aquifer which incorporates actual water table data from several locations, soil properties according to depth and including extraction wells and recharge sources also, thus requiring matching conditions depending on the depth; setting the way for the end purpose of developing and alternative computational toolkit for the analysis of groundwater flow, this will allow us to predict the behaviour of the system and to give policies for the sustainable exploitation and management of the aquifer.

\section*{Acknowledgments}

We thank VIEP-BUAP and DGDI-BUAP for their financial support during the realization of this work, also we thank PRODEP-SEP for supporting publication expenses. We also thank Javier M. Hernandez-Lopez for his useful insight and comments in the subject of this paper.

\section*{Conflict of Interest}

The author(s) declare(s) that there is no conflict of interest regarding the publication of this paper.

\section*{Data Availability}

The hydrogeological data supporting this study are from previously reported studies and datasets, which have been cited.


\end{document}